\documentclass[aps,prl,twocolumn,10pt,superscriptaddress]{revtex4-2}

\usepackage{graphicx}
\usepackage[utf8]{inputenc}
\usepackage[T1]{fontenc}
\usepackage{amssymb}
\usepackage{amsmath}
\usepackage[colorlinks=true,linkcolor=blue,citecolor=blue]{hyperref}
\usepackage[capitalize]{cleveref}
\usepackage{physics}

\renewcommand{\d}{\mathrm{d}}
\newcommand{\cond}{\mathrm{c}}
\newcommand{\gammadamp}{\gamma_p}
\newcommand{\mysection}[1]{\textit{#1.---}}

\begin{document}
\title{Quantum tunneling of a measurement-driven active particle} 

\author{Tobias Nadolny}
\affiliation{Department of Physics, University of Basel, Klingelbergstrasse 82, 4056 Basel, Switzerland}
\affiliation{Institute for Theoretical Physics, ETH Zürich, 8093 Zürich, Switzerland}
\author{Matteo Brunelli}
\affiliation{JEIP, UAR 3573 CNRS, Coll\`ege de France, PSL Research University, 11 Place
Marcelin Berthelot, 75321 Paris Cedex 05, France}
\author{Christoph Bruder}
\affiliation{Department of Physics, University of Basel, Klingelbergstrasse 82, 4056 Basel, Switzerland}
\author{Michel Fruchart}
\affiliation{Gulliver, ESPCI Paris, Université PSL, CNRS, 75005 Paris, France}

\date{November 12, 2026}
\begin{abstract}
Active particles can exhibit persistent motion set by their orientation.
We demonstrate this key property for a quantum active particle: The particle's motion is locked to the orientation of its spin due to spin-orbit coupling and damping, and the orientation is made persistent through continuous measurement via the quantum Zeno effect.
As a result, individual quantum trajectories feature an active run-and-tumble motion.
When the particle impinges on a potential barrier, its persistent velocity leads to accumulation near the barrier, in contrast with the reflection of a free quantum particle.
We show that quantum tunneling enhances the active particle's transmission rate compared to that of an analogous classical particle.
\end{abstract}

\maketitle
Active particles are nonequilibrium entities that use local energy sources to exert forces~\cite{Marchetti_2013,Bechinger_2016,teVrugt_2025b}.
Interacting active particles exhibit phenomena ranging from flocking~\cite{Toner_Tu_1995,Vicsek_1995} and motility-induced phase separation~\cite{Cates_Tailleur_2015} to nonreciprocal phase transitions towards time-dependent states~\cite{Fruchart_2021,Nadolny_2025}.
A key feature of classical active particles is their persistent, self-propelled motion, which gives rise to characteristic phenomena at the level of individual trajectories, such as accumulation near boundaries~\cite{Bechinger_2016}.
Quantum versions of active particles have recently been considered in various models and platforms~\cite{Khasseh_2025,Penner_2025,Antonov_2025,antonov2026,gipouloux2026,steiner2026,lee2026,Romano_2026,Brunner_2026,Radhakrishnan_2026,Burgardt_2026,Assent2026}.
Yet, it remains an open question how persistent motion manifests itself in quantum systems.
What happens, in particular, when a quantum active particle encounters a potential barrier?
Unlike classical particles, quantum particles can tunnel through a barrier.
Does a quantum active particle accumulate near the barrier, does it tunnel through it, and what is the interplay between these processes?

\begin{figure}
    \centering
    \includegraphics[width=3.4in]{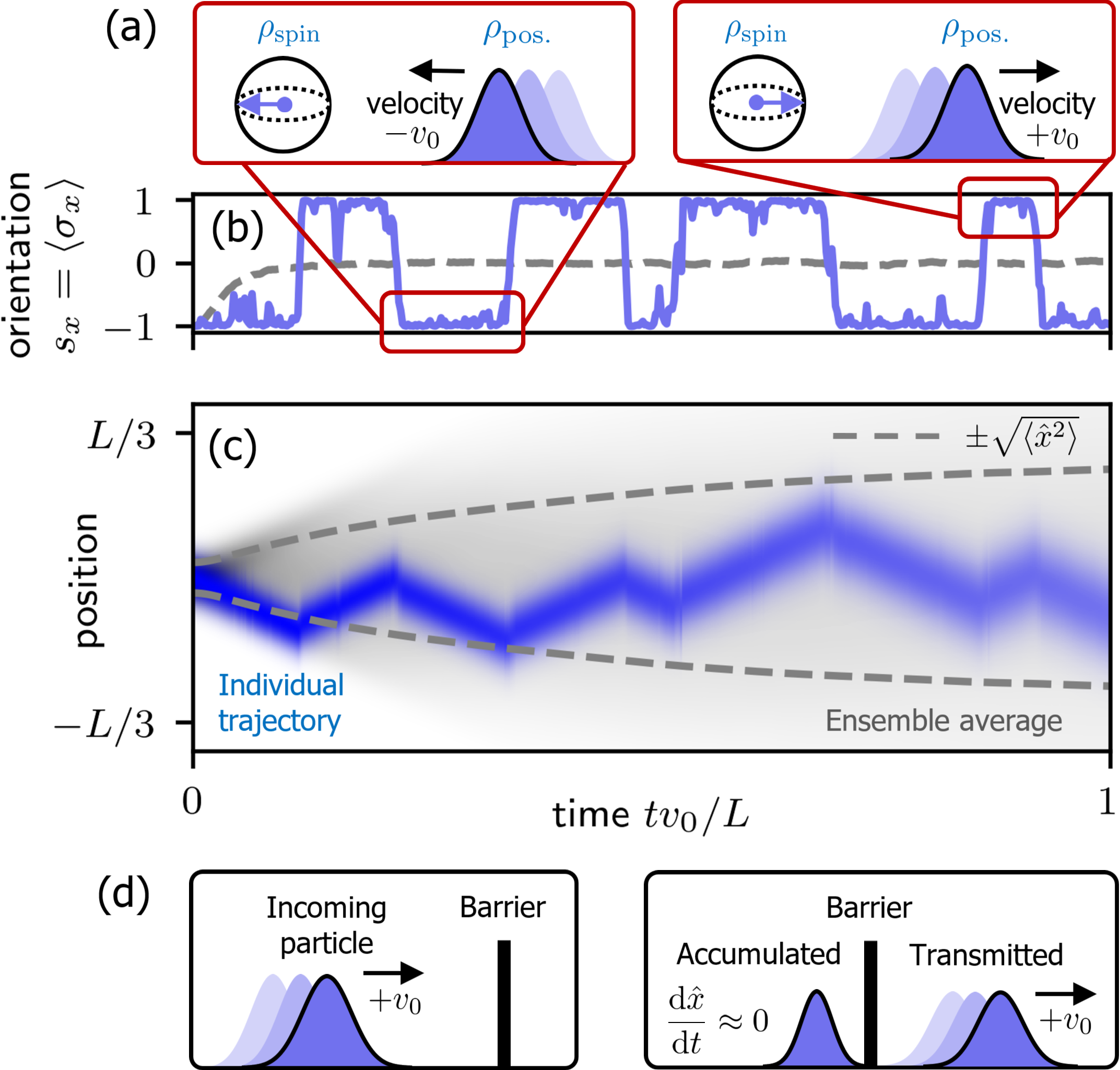}
    \caption{
    (a)
    Spin-dependent motion.
    The particle's positional distribution is depicted as a Gaussian;
    increasing opacity indicates passage of time.
    On the left side, the particle's spin degree of freedom is close to the minus $\sigma_x$ eigenstate, and thus the particle moves with velocity $-v_0$.
    On the right side, we depict a plus $\sigma_x$ eigenstate and velocity $+v_0$.
    (b)
    Orientation dynamics of a single quantum trajectory (blue) and the ensemble average (gray).
    (c)
    Position dynamics.
    In blue, we show the positional distribution corresponding to the trajectory in (b): If $s_x \approx +1$, the particle moves up, and if $s_x \approx -1$, the particle moves down.
    The gray scale shows the ensemble average, and the gray dashed line shows the root-mean-square displacement.
    (d)
    Active quantum tunneling.
    The particle with initial velocity $+v_0$ impinges on a barrier.
    Assuming no spin flip, a fraction of its wavefunction accumulates at the barrier and stays at rest.
    The other fraction is transmitted through the barrier and continues at velocity~$+v_0$.
    }
    \label{fig:fig1}
\end{figure}

To address these questions, we introduce a minimal model of a quantum active particle in which the orientation of the particle is encoded in a quantum spin.
This spin effectively controls the velocity of the particle through a combination of spin-orbit coupling and momentum damping; see \cref{fig:fig1}(a).
The spin itself is continuously monitored, enforcing persistence in its dynamics akin to the quantum Zeno effect~\cite{Misra_1977,Facchi_2008,Greenfield_2025}, which in turn results in persistent motion.
We demonstrate that the resulting quantum active particles undergo a run-and-tumble motion in which the velocity is locked to the measured spin; see \cref{fig:fig1}(b,c).
In contrast with existing works on quantum active particles, which focus on average properties, our approach based on quantum trajectories makes self-propulsion apparent through continuous measurement~\citep{Wiseman_2009}; compare the probability densities in blue (individual trajectory) and gray (average) in \cref{fig:fig1}(c).
We show that the persistent motion of our quantum active particles leads to their accumulation near a barrier, offering them repeated opportunities to tunnel; see \cref{fig:fig1}(d).
As a consequence, quantum active particles have an increased escape rate with respect to both classical active and quantum passive particles.

\mysection{Spin-dependent motion}
The quantum active particle discussed in our work moves in one dimension, and it carries an internal quantum degree of freedom representing the particle's orientation, which we choose to be a spin-1/2; see \cref{fig:fig1}(a).
The particle's dynamics is described by the Hamiltonian
\begin{equation}
    H = 
    { \frac{\hat p^2}{2m} + V(\hat x) }
    +
    { H_{\text{spin}}}
    +
    { v_0 \hat{p} \sigma_x} \, ,
    \label{eq:Hm}
\end{equation}
where $\hat{x}$ and $\hat{p}$ are the position and momentum operators, the spin is described by Pauli matrices $\sigma_{x,y,z}$, and $m$ is the particle's mass. 
The first and second terms correspond to the kinetic and potential energy of the particle, respectively.
The Hamiltonian $H_{\text{spin}}$ describing the spin dynamics will be specified later.
Crucially, the last term, where $v_0$ is a characteristic velocity, describes a coupling between the motion and the spin.

The coupling between spin and motion in Eq.~\eqref{eq:Hm} is called Rashba spin-orbit coupling~\cite{Winkler_2003}; it is well known in spintronics~\cite{Zutic2004} and topological insulators~\cite{HasanKane_2010}.
It can also be realized synthetically in Bose-Einstein condensates~\cite{Lin_2011,Zhai_2015} or in trapped ions~\cite{Haljan_2005,Gerritsma_2010}.
Here, we consider this coupling as a way to make the active particle move in the direction given by the spin.
This becomes apparent via the quantum Heisenberg equation of motion
\begin{equation}
    \frac{\d \hat{x}}{\d t} = \frac{\hat{p}}{m}+ v_0 \sigma_x \, ,
    \label{eq:eom}
\end{equation}
which shows that, in addition to its motion due to inertia, the particle moves at velocity $v_0$ in the direction set by its spin $\sigma_x$.
A similar Hamiltonian coupling between motion and orientation has been discussed for classical self-propelled particles~\cite{Bore_2016,Casiulis_2020,Casiulis2020b,Tasaki2020,Casiulis2026,Bhattacharya2026,Loewe2018}, for quantum active particles~\cite{Penner_2025,gipouloux2026} and in the context of quantum random walks~\cite{Aslangul_2005,Zhang_2010,Attal2012, Bauer_2013,Bauer2014, Sinayskiy_2019}.

\mysection{Momentum damping}
Classical active particles are generally subject to dissipation, such as fluid friction, which damps their momentum and tends to suppress inertial effects~\cite{Lowen2020}. 
We introduce momentum damping by coupling the particle’s position to a collection of harmonic oscillators, which constitute a bath described by the Caldeira-Leggett model~\cite{Caldeira1983,Breuer}.
As shown in the End Matter, this leads to the Heisenberg equation of motion
\begin{equation}
    \frac{\d \hat{p}}{\d t} = - \gammadamp \hat{p} - V'(\hat{x}) \, ,
    \label{eq:eom2}
\end{equation}
where $\gammadamp$ is the rate of damping.
This bath damps the canonical momentum rather than the kinetic momentum.
Therefore, the bath suppresses inertial motion, so that Eq.~\eqref{eq:eom} yields a velocity $\d\hat{x}/\d t \simeq v_0 \sigma_x$ controlled entirely by the spin orientation.
Hence, Eqs.~(\ref{eq:eom}--\ref{eq:eom2}) imply that the position dynamics are governed by the spin dynamics.
This behavior is illustrated in \cref{fig:fig1}(a).
A similar interplay between momentum damping and spin-orbit coupling has been considered in the context of chirality-induced spin selectivity \cite{Volosniev2021,Ghazaryan2023}.

\begin{figure*}
    \centering
    \includegraphics[width=\linewidth]{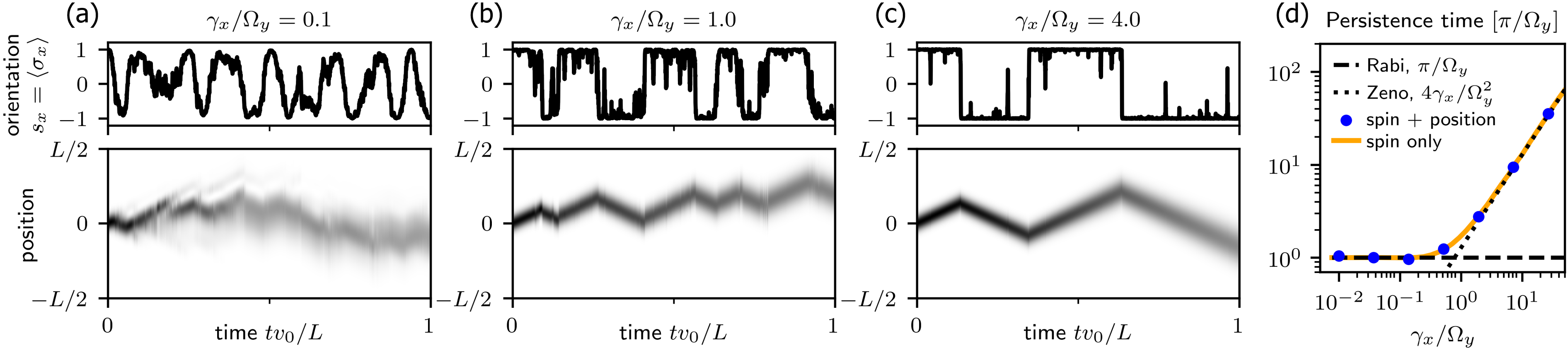}
    \caption{
    Quantum run-and-tumble motion.
    (a-c)
    Individual quantum trajectories for three different measurement rates (no momentum damping).
    The top panel shows the orientation; the bottom panel shows the particle's positional probability density on a linear gray scale.
    (d)
    Persistence time.
    The orange line is obtained from the spin-only dynamics, which agrees well with the full simulation including position (blue dots).
    The dashed and dotted lines indicate the Rabi ($\gamma_x \ll \Omega_y$) and Zeno limits ($\gamma_x \gg \Omega_y$). 
    Numerical simulations use a discretized version of \cref{eq:sse} on a lattice with spacing $a =1/(mv_0)$ and $L/a = 200$ lattice sites,
    $\Omega_y = 50v_0/L$; see End Matter for details.
    The initial state is the plus $\sigma_x$ eigenstate and a Gaussian positional distribution centered at zero with standard deviation $7.1a$.
    }
    \label{fig:quantum_run_tumble}
\end{figure*}

\mysection{Activity through measurements}
As the velocity of the quantum particle is set by its spin state, we can influence the particle's motion by acting on the spin.
We do so through continuous measurements, which directly affect the state of quantum systems through measurement backaction.
The backaction associated with the continuous monitoring of an observable $\mathcal{O}$ can be modeled by a stochastic version of the Schr\"odinger equation~\cite[Eq.~(4.76)]{Wiseman_2009}
\begin{equation}
i\hbar \frac{\d\ket{\psi}}{\d t} =  \left[H -i \hbar \frac{\gamma_x}{2} \tilde{\mathcal{O}}^2 \right] \ket{\psi} + i\hbar \sqrt{\gamma_x} \tilde{\mathcal{O}}  \xi(t) \ket{\psi} \, ,
 \label{eq:sse}
\end{equation}
in which $\tilde{\mathcal{O}} = \mathcal{O} - \langle \mathcal{O} \rangle$,
and $\gamma_x$ describes the strength of the measurement (we will later choose $\mathcal{O}=\sigma_x$, hence the subscript $x$).
Here, $\ket{\psi}$ is the state of both spin and position. 
We focus on spin monitoring and leave positional measurements and feedback for future work.
Monitoring produces a stochastic measurement record, whose random part is the Gaussian white noise $\xi(t)$ (with zero mean and unit variance).
The noise acts back on the state via the last term of \cref{eq:sse}, giving rise to a stochastic evolution known as a quantum trajectory; see \cref{fig:fig1}(b).
Through the spin-orbit coupling, these stochastic changes in the spin orientation translate into changes in the particle’s velocity as shown in \cref{fig:fig1}(c), and hence the backaction on the spin also affects the spatial part.

The momentum-damping bath producing the $\gamma_{p}$ term in Eq.~\eqref{eq:eom2} is an unmonitored environment, which makes the conditional state of the particle generally mixed, even when conditioned on the measurement record.
The corresponding description is given in terms of the conditional density operator $\rho_\cond(t)$ for the combined state of position and spin of the particle, generalizing the pure state entering Eq.~\eqref{eq:sse}.

Our full model of an active quantum particle is defined by the stochastic master equation (from now on, we set $\hbar = 1$)    
\begin{equation}
    \begin{split}
    \frac{\d {\rho}_\cond}{\d t} = &-i[H,\rho_\cond]
    + 
    \mathcal{L}_{\text{damp}}
    \rho_\cond+
    \mathcal{L}_{\text{meas}}
    (\xi)\rho_\cond \, .
    \end{split}
    \label{eq:sme}
\end{equation}
In addition to the coherent dynamics generated by the Hamiltonian of \cref{eq:Hm}, which gives rise to the Heisenberg equation~\eqref{eq:eom}, it includes two dissipative terms:
The superoperator $\mathcal{L}_{\text{damp}}$ describes the momentum-damping bath responsible for the damping in \cref{eq:eom2}, while $\mathcal{L}_{\text{meas}}$ accounts for the measurement; see End Matter for explicit expressions.
    We emphasize that $\rho_\cond$ depends on the noise realization, so each trajectory is different.
As we now show, continuous monitoring of a spin observable can lead to active persistent motion.

\mysection{Quantum run-and-tumble}
We illustrate the active motion by showing a quantum run-and-tumble behavior.
Run-and-tumble motion is an idealized version of the motion of bacteria, composed of time intervals called runs where a particle moves in a constant direction, separated by rapid changes of direction called tumbles~\cite{Schnitzer1993,Berg2004,Cates_Tailleur_2015}.
To mimic this behavior, we set the spin Hamiltonian to $H_{\text{spin}}=\Omega_y\sigma_y/2$, which induces oscillations of the spin orientation around the $\sigma_y$ axis.
The combination of these spin dynamics with spin-orbit coupling has been discussed in the context of Zitterbewegung, where the particle spreads ballistically even in the long-time limit~\cite{Gerritsma_2010,LeBlanc_2013}.
Here, we include a continuous measurement of the observable $\mathcal{O}=\sigma_x$.
This type of measurement continuously monitors the $\sigma_x$ value of the spin and thereby projects it onto the $\sigma_x$ basis; it is employed, for instance, in the dispersive readout of superconducting qubits~\cite{Gambetta_2008}.
In a Bose-Einstein condensate, it may be engineered via Faraday interactions and polarization measurements~\cite{Hammerer_2010}.
For now, we set $V(\hat x) = 0$ so that $\hat{p}$ is conserved and, by picking an initial state with $\expval{\hat p}=0$, no momentum damping is needed for the spin-controlled velocity.
The state $\rho_\cond$ remains pure in this case, which significantly reduces the computational complexity, because we can use the stochastic Schr\"odinger equation \eqref{eq:sse} instead of the stochastic master equation~\eqref{eq:sme}.

\Cref{fig:quantum_run_tumble} shows individual quantum trajectories obtained from \cref{eq:sse} with $\mathcal{O}=\sigma_x$.
The top row shows the continuously measured orientation $s_x(t) \equiv \langle\sigma_x\rangle$, while the bottom row shows the associated spatial probability distribution $|\psi(x)|^2$.
The particle indeed follows the orientation as expected from \cref{eq:eom}: It moves towards positive $x$ when $s_x$ is positive, and towards negative $x$ when $s_x$ is negative; compare top and bottom rows of \cref{fig:quantum_run_tumble}(a-c).

For small measurement rates, the orientation $s_x$ undergoes noisy oscillations with a frequency close to $\Omega_y$ [\cref{fig:quantum_run_tumble}(a, top)], similar to the Rabi oscillations that would occur with no measurements.
During the oscillations, the spin passes through superposition states of the two $\sigma_x$ eigenstates.
In this case, the particle's positional distribution can split into components moving in opposite directions; see \cref{fig:quantum_run_tumble}(a, bottom), where this is most visible in the first half. 

As the measurement rate increases, the spin tends to be projected to a pure state where $s_x = \pm 1$ [\cref{fig:quantum_run_tumble}(b,c, top)].
This is known as the quantum Zeno effect~\cite{Misra_1977,Facchi_2008,Greenfield_2025}.
As we show in the End Matter, in the Zeno limit $\gamma_x \gg \Omega_y$, the orientation follows~\cite{Gambetta_2008}
\begin{equation}
     \frac{\d {s}_x}{\d t} \simeq - 2\Gamma s_x + 2 \sqrt{\gamma_x}(1-s_x^2)\xi(t)\, ,
     \label{eq:gammaswitch}
\end{equation}
where $\Gamma \equiv \Omega_y^2/4\gamma_x$.
Equation~\eqref{eq:gammaswitch} produces a telegraph-like process, where $s_x(t)$ jumps between $\pm 1$ (also described in Ref.~\cite{kundu2025runtumbledynamicsbiased} in analogy with active matter).
As a result of the telegraph signal in the orientation, the quantum trajectory of the particle consists of periods with constant velocity resembling a classical run-and-tumble motion.
The persistence time, defined as the average time between a sign flip in the orientation $s_x$ is shown in
Fig.~\ref{fig:quantum_run_tumble}(d), which compares limiting behaviors obtained from the spin-only equations with the persistence times measured in numerical simulations of the stochastic Schrödinger equation \eqref{eq:sse} including space.

Our first key result is that the persistent motion of our quantum active particle is directly visible in individual quantum trajectories; see \cref{fig:quantum_run_tumble}(b,c, bottom).
The ensemble average shown in \cref{fig:fig1}(c) via the gray scale captures this behavior only indirectly via the crossover of the mean-square displacement from ballistic to diffusive (gray dashed line).
The run-and-tumble behavior observed in individual quantum trajectories is reminiscent of the open quantum random walk of Refs.~\cite{Bauer_2013,Bauer2014} where the internal state of the walker influences how it changes its position.
By formulating our model in the continuum and including inertia, we can study the behavior of the particle's quantum wavefunction in the presence of potential barriers.

\begin{figure}
    \centering
    \includegraphics[width = \linewidth]{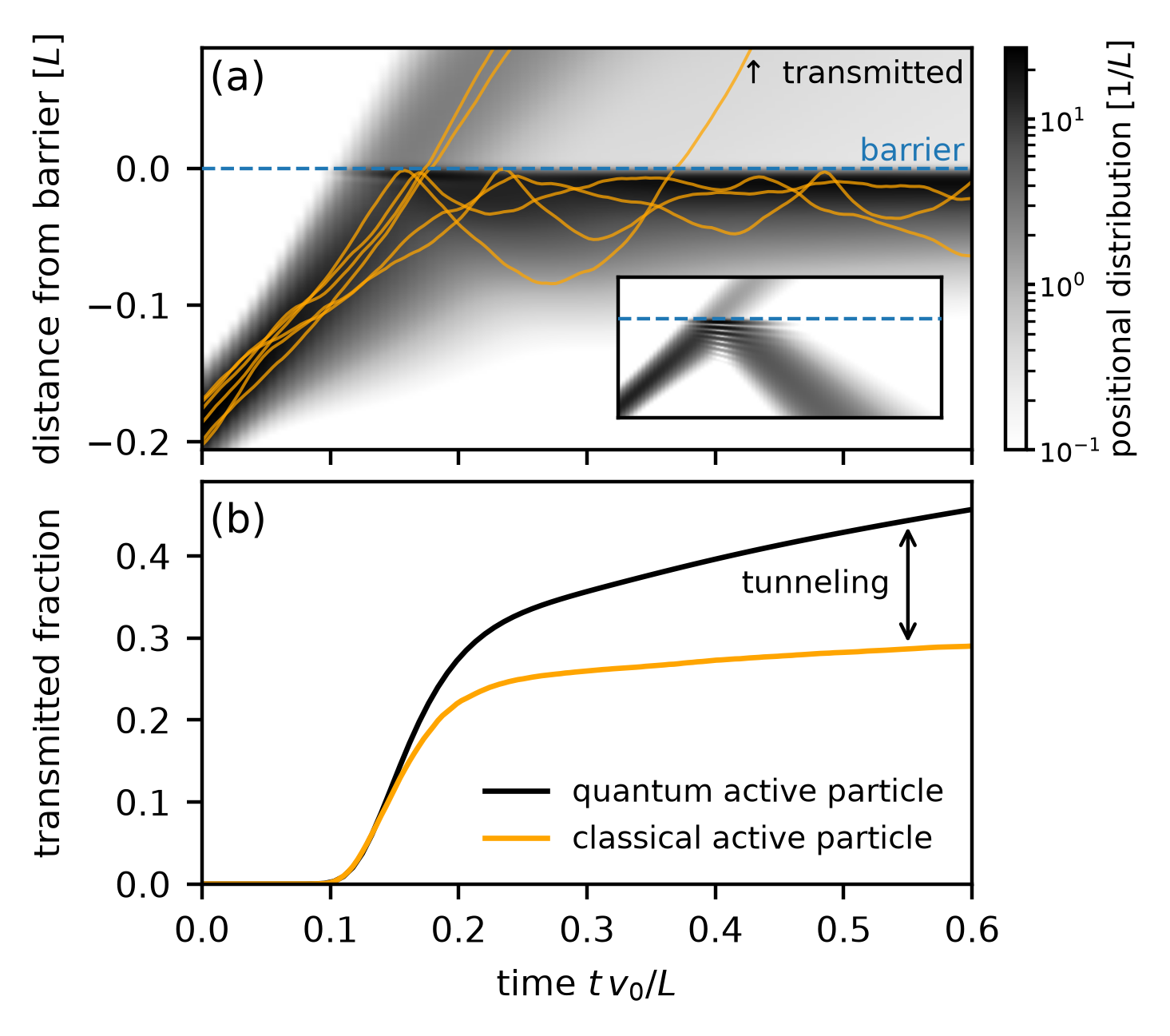}
    \caption{
    Active quantum tunneling.
    (a)
    Time evolution of the probability distribution (gray scale, logarithmic) of the quantum active particle with $\sigma_x \rightarrow +1$ (no-tumble limit) impinging on a Gaussian barrier at $x=0$.
    The orange lines show samples of an analogous ensemble of classical particles.
    Inset: standard tunneling of a passive particle ($\gammadamp=0$ and $v_0=0$) with the same initial velocity.
    (b)
    Transmitted fraction (probability at $x>0$) of the quantum particle (black line) and the classical particle (orange line, averaged over 40,000 realizations).
    Parameters:
    We simulate \cref{eq:sme} on a lattice with spacing $a = 1/(4mv_0)$, and $L/a=1088$ sites, setting $\gammadamp = 0.04mv_0^2 = T/5$.
    Initially, the quantum particle is in a pure Gaussian state of width $16a$.
    Gaussian barrier: $V(\hat x)=h\,\exp(-\hat x^2/2w^2)$ with $w=3a$, and height $h=mv_0^2$.
    }
\vspace{-4pt}
    \label{fig:tunneling}
\end{figure}

\mysection{Active quantum tunneling}
To analyze the role of quantum tunneling in active matter, we now study the behavior of the quantum active particle impinging on a potential barrier, as illustrated in \cref{fig:fig1}(d).
Tunneling of particles with spin-orbit coupling has been considered in a Bose-Einstein condensate~\cite{Zhang_2012,Qin_2019} and for trapped ions~\cite{Gerritsma_2011}, but without momentum damping or measurement.
We choose a Gaussian-shaped potential barrier $V(\hat x)$ in the Hamiltonian of \cref{eq:Hm}.
As momentum is no longer conserved, we now include the momentum-damping Caldeira-Leggett bath.
We focus on an individual trajectory in the strong-measurement limit, in which the spin state remains in one of the pure eigenstates of $\sigma_x$, say the plus $\sigma_x$ eigenstate.
We therefore solve the positional part of \cref{eq:sme} with the replacement $\sigma_x \rightarrow +1$ to impose a fixed spin.
In this limit, the system is equivalent to a passive particle (i.e., $v_0 =0$) in a tilted potential $V\rightarrow V - \hat{x}v_0 m\gammadamp$ via the transformation $\hat p \rightarrow \hat p +mv_0$.
This equivalence only works in the no-tumble limit and breaks down for the run-and-tumble motion.

\Cref{fig:tunneling}(a) shows the positional probability density of the quantum active particle impinging on the potential barrier (gray scale).
The results are compared with the dynamics of a classical active particle which are obtained by converting \cref{eq:sme} into the corresponding classical Langevin dynamics (orange lines;
see End Matter for details).
The quantum particle and the ensemble of classical particles start with the same phase-space distribution and they are subject to the same bath.
In both cases, a fraction of the probability density is reflected by the barrier, while another fraction passes through.
While the classical active particle can only escape by thermal excitation over the barrier, the quantum active particle can also tunnel.
As \cref{fig:tunneling}(b) shows, the quantum transmission (black line) is therefore higher than the classical one (orange line).

The difference between passive and active particles becomes most clear when considering the reflected fraction.
The passive particle moves away from the barrier after reflection because momentum is reversed [see inset of \cref{fig:tunneling}(a)]. 
The active particle is also initially reflected, but it then tends to resume its persistent motion in the direction of its internal orientation, which still points in the direction of the barrier.
As a result, the reflected part stays close to the potential barrier after the initial impact.
The particle ends up in a metastable state at a finite average distance from the barrier [see \cref{fig:tunneling}(a)] where it keeps on being transmitted through the barrier.
The resulting cumulative transmission continues to increase.
The transmission rate of the quantum active particle is higher than that of both a classical active particle due to tunneling [compare the slopes in \cref{fig:tunneling}(b)] and a passive quantum particle, which is reflected and for which the asymptotic slope vanishes (not shown).
The demonstration of the accumulation, tunneling, and enhanced escape rate of a quantum active particle is the second key result of our work.

\mysection{Conclusions}
We have introduced a model of a quantum active particle based on
(i) Hamiltonian spin-momentum coupling, (ii) dissipative damping of the canonical momentum, and (iii) continuous measurements.
We have shown how active motion arises from these three ingredients:
The first two yield a velocity controlled by the spin, while the measurement influences the state of the spin and thereby makes the motion persistent.
Individual quantum trajectories, as opposed to ensemble averages, make the persistent motion directly visible.
Further, we have shown how the interplay between quantum tunneling and boundary accumulation of self-propelled particles leads to an enhanced transmission rate compared to that of both passive quantum particles and classical active particles.

Looking ahead, one may extend the control over the particle's orientation by measurement backaction to include feedback, enabling the exploration of quantum counterparts of informational active matter~\cite{VanSaders_2026}.
In addition, our work can be generalized to higher spatial dimensions as well as to interacting particles, which may align their orientations to form a flock of quantum particles.

\begin{acknowledgments}
\mysection{Acknowledgments}
We would like to thank
M.~Bukov,
R.\ Fazio,
M.\ Heyl,
R.\ Moessner,
F.\ von Oppen,
M.\ Schirò,
T.\,L.\ Schmidt,
and
M.\ te Vrugt
for stimulating discussions. 
T.\,N.\ and C.\,B.\ acknowledge financial support from the Swiss National Science Foundation individual grant (No.\ 200020 200481).
M.\,B.\ acknowledges 
funding from the European Research Council (ERC) under the European Union's Horizon 2020 research and innovation programme (Grant agreement No.\ 101002955 - CONQUER).
\end{acknowledgments}

\clearpage
\section{End Matter}
\appendix
\mysection{Full model}
The stochastic (Itô) quantum master equation for the conditional density operator $\rho_\cond$ of position and spin reads
\begin{equation}
    \begin{split}
    \dot{\rho}_\cond = &-i[H,\rho_\cond]
    + 
    \mathcal{L}_{\text{damp}}\rho_\cond+
    \mathcal{L}_{\text{meas}}(\xi)\rho_\cond \, .
    \end{split}
    \label{eq:master_endmatter}
\end{equation}
Here, $H$ is the Hamiltonian given by Eq.~\eqref{eq:Hm}.
Using $\mathcal{D}[o]\rho_\cond = o\rho_\cond o^\dag - (o^\dag o \rho_\cond + \rho_\cond o^\dag o)/2$ and introducing the expectation value $\expval{o} \equiv \Tr[\rho_\cond o]$, the spin measurement term is 
\begin{equation}
     \mathcal{L}_\text{meas}(\xi)
     \rho_\cond=\gamma_x\mathcal{D}[\sigma_x]\rho_\cond + \sqrt{\gamma_x}\big(
     \sigma_x\rho_\cond+\rho_\cond \sigma_x
     - 2\langle \sigma_x\rangle\rho_\cond\big)\,\xi \, ,
\end{equation}
which is equivalent to the measurement in the stochastic Schr\"odinger equation \eqref{eq:sse}.
The Caldeira-Leggett bath is
\begin{equation}
    \mathcal{L}_{\text{damp}}\rho_\cond =
    - \frac{i\gammadamp}{2}\,\big[\hat x,\{\hat p,\rho_\cond\}\big]
    -{m\gammadamp T}\,\big[\hat x,[\hat x,\rho_\cond]\big]\, ,
\end{equation}
where $\gammadamp$ is the relaxation rate of the canonical momentum $\hat p$ and $T$ is the bath temperature ($k_\mathrm{B}=1$).
Equation~\eqref{eq:eom2} follows using $\Tr[\hat p \mathcal{L}_{\text{damp}}\rho_\cond] = -\gammadamp \Tr[\hat p \rho_\cond]$.
Even though the master equation is not of Lindblad form, it is valid in the regime $T>\gammadamp$, where the density operator stays nearly positive (in our simulations, the negative eigenvalues of the density operator stay above $-4\times 10^{-4}$).

\vspace{1em}
\mysection{Spin dynamics}
The averages $s_i = \langle \sigma_i \rangle$ of the spin components follow the stochastic Bloch equations, derived from \cref{eq:master_endmatter},
\begin{equation}
    \begin{split}
    \dot s_x &= +\Omega_y s_z + 2\sqrt{\gamma_x}(1-s_x^2)\xi \, ,\\
    \dot s_y &= - 2\gamma_x s_y - 2\sqrt{\gamma_x} s_x s_y \xi -2v_0\expval{\sigma_z \hat p} \, ,\\
    \dot s_z &= -\Omega_y s_x - 2\gamma_x s_z - 2\sqrt{\gamma_x} s_x s_z \xi +2v_0\expval{\sigma_y \hat p}\, ,
    \end{split}
    \label{eq:sxyz}
\end{equation}
These equations are in general not closed because of the $\langle \sigma_{y,z} \hat{p}\rangle$ terms.
In the Zeno limit, these terms factorize into $s_{y,z} \expval{\hat p}$ as the strong measurement forces the spin into a pure state.
For the quantum run-and-tumble motion, momentum is conserved and we choose $\expval{\hat{p}} = 0$, so that we can neglect these terms.
\Cref{fig:quantum_run_tumble}(d) shows a good agreement between the spin-only dynamics obtained from \cref{eq:sxyz} with $v_0 = 0$ (orange line) and the full simulation including position obtained from \cref{eq:sse} with $\mathcal{O} = \sigma_x$

Equations~\eqref{eq:sxyz}, closed by setting $\expval{\sigma_{z,y}\hat p} =0$, qualitatively capture the behavior of $s_x$ in the top row of Fig.~\ref{fig:quantum_run_tumble}.
In the absence of the measurement, the spin performs Rabi oscillations around $\sigma_y$: $\dot{s}_x = +\Omega_y s_z$ and $\dot{s}_z = -\Omega_y s_x$.
A weak measurement introduces fluctuations around the oscillations.
In the Zeno limit $\gamma_x \gg \Omega_y$, we can set $\dot s_z = 0$ and find \cref{eq:gammaswitch} by adiabatic elimination~\cite{Gambetta_2008}.
This equation describes a telegraph-like process [see \cref{fig:quantum_run_tumble}(c, top)], which, for $\gamma_x \gg \Omega_y$, can be cast as
\begin{equation}
        \d s_x = -2s_x \d N(t) \, ,
        \label{eq:s_x_telegraph2}
\end{equation}
where $\mathrm{d}N$ is a Poisson point process with expectation value $\mathbb{E}[\mathrm{d}N(t)] = \Gamma \d t$.

\vspace{1em}
\mysection{Classical limit}
The Caldeira-Leggett model allows for a direct comparison of the quantum dynamics with the corresponding classical dynamics at the same temperature.
In the Zeno limit, we can replace the spin with its expectation value $s_x$ that follows \cref{eq:s_x_telegraph2}.
The classical dynamics equivalent to the master equation~\eqref{eq:master_endmatter} are
\begin{equation}
    \begin{split}
        \dot x &= p/m + v_0 s_x \,,\\
        \dot p &= -V'(x) - \gammadamp p + \sqrt{2m\gammadamp T}\zeta \, ,
    \end{split}
    \label{eq:Langevin}
\end{equation}
where $\zeta$ is a Gaussian white noise with zero mean and unit variance.
The classical analog is obtained by transforming \cref{eq:master_endmatter} to the time evolution of the Wigner function and dropping third and higher order derivatives~\cite[Eq.~(3.13)]{Zurek_2003}.
Equations~\eqref{eq:Langevin} are used to obtain the classical trajectories in \cref{fig:tunneling}(a).

In the overdamped limit, the dynamics governed by \cref{eq:Langevin} become
\begin{equation}
    \dot x = v_0 s_x - V'(x)/(m\gammadamp) + \sqrt{D}\zeta \, ,
\end{equation}
where $s_x$ follows the telegraph signal specified in \cref{eq:s_x_telegraph2} and $D=2T/(m\gammadamp)$.
See Ref.~\cite{Malakar_2018} for more details on run-and-tumble motion.
In the absence of a potential, the particle displays the characteristic run-and-tumble motion.
At a potential barrier, the particle stops when $m\gammadamp v_0 = V'(x)$, unless fluctuations cause it to pass the barrier; see \cref{fig:tunneling}.

\vspace{1em}
\mysection{Numerical implementation}
\label{app:lattice}
To simulate the dynamics, we discretize real space into $N$ lattice sites separated by a lattice spacing $a$.
The state of the particle and its spin is represented by amplitudes on each lattice site $j \in \{1,...,N\}$ and for each spin component.
On the lattice, the kinetic and spin-orbit terms of the Hamiltonian in \cref{eq:Hm} become nearest-neighbor hopping terms with real and imaginary amplitudes $\eta$ and $i\eta_s$, respectively,
\begin{equation}
\begin{split}
    H_\mathrm{lat} = \sum_{j} \Big[
    &-\hat{\mathbf{c}}^\dagger_{j+1}(\eta - i\eta_s\sigma_x)\,\hat{\mathbf{c}}_j + \mathrm{H.c.} \\
    &+ \hat{\mathbf{c}}^\dagger_{j}\big({\Omega_y}\sigma_y/2 + V_j\big)\hat{\mathbf{c}}_{j}\Big] .
\end{split}
\label{eq:H_lat}
\end{equation}
Here, $\hat{\mathbf{c}}_j = (\hat{c}_{j,\uparrow},\hat{c}_{j,\downarrow})$ annihilates the particle with spin $\uparrow$ or $\downarrow$ on site $j$, 
$\eta = 1/(2 m a^2)$ is the nearest-neighbor hopping rate, and the spin-dependent hopping rate is $\eta_s = v_0/(2a)$. 
We take $V_j$ to form a Gaussian barrier.
In the limit $a \to 0$ at fixed $Na$, Eq.~\eqref{eq:H_lat} converges to the continuum Hamiltonian of~\cref{eq:Hm}.

In the context of the tunneling [see \cref{fig:tunneling}], we focus on the strong-measurement limit, which enforces the spin state to remain in one of the $\sigma_x$ eigenstates.
Therefore, we simulate only the spatial part of the lattice Hamiltonian, setting $\sigma_x \rightarrow +1$, $\Omega_y = 0$, and $\hat{\mathbf{c}}_j \rightarrow \hat{{c}}_j$, which annihilates the particle (in the plus $\sigma_x$ eigenstate) at site~$j$.

\end{document}